\documentstyle[twocolumn,aps]{revtex}
\input psfig.tex
\begin{document}
\draft
\title{\bf Bose-Einstein Condensation in a Trap: the Case of a 
Dense Condensate}
\author{Klaus Ziegler and Alok Shukla}
\address{Max-Planck-Institut f\"ur Physik Komplexer Systeme,
Au\ss enstelle Stuttgart, Postfach 800665, D-70506 Stuttgart,
Germany}
\date{\today}
\maketitle 
\begin{abstract}
We consider the Bose-Einstein condensation of atoms in a trap where the
density of particles is so high that the low density approach of Gross and
Pitaevskii will not be applicable. For this purpose we use the slave boson
representation which is valid for hard-core bosons at any density. This
description leads to the same results as the Gross-Pitaevskii approach
in the low density limit, but for higher densities, it predicts the
depletion of the
condensate in the regions where the density
of the atomic cloud is high.
\end{abstract}
\section{Introduction}
Atoms in a magnetic trap present an interesting system for the analysis of
highly degenerate gases. The trap plays the role of a 
three-dimensional confining potential well  and by using advanced experimental 
techniques such as laser and evaporative cooling, it is possible to study the 
the gas over a wide range of parameters like temperature or density of
particles. One of the most spectacular achievements of such techniques 
was the observation of the Bose-Einstein (BE) condensation in 
gases composed of alkali atoms~\cite{A}. BE condensation,
originally studied in terms of the ideal (non-interacting) Bose gas
requires a  minimal density of bosons
$\rho_0=g_{3/2}(1)/\lambda^3$, where $\lambda=(2\pi\hbar^2/mk_BT)^{1/2}$
is the de Broglie wave length and $g_{3/2}(1)\approx 2.612$.
On the other hand, it is known that the density of the condensate in an
interacting Bose gas at high density is depleted 
if the total density exceeds a certain value. For instance, the condensate
in the bulk of $^4$He is only about 10\% at zero temperature.
On the surface, however,
the condensate can reach almost 100\% because of the reduced total density.
This phenomenon was observed in numerical simulations of an interacting
Bose gas \cite{LRP} and in analytic calculations including an attractive
interaction\cite{GS} or in a slave boson approach to a hard-core Bose gas
\cite{Z}. 
The effect can be understood as a reduction of long-range correlations,
necessary for the formation of a condensate, which is
caused by increasing fluctuations due to an increasing density of
interacting particles.
There is an approach to the dilute interacting Bose gas due to Ginzburg and
Pitaevskii \cite{P} and Gross \cite{G}, analogous to the Ginzburg-Landau
approach for second order phase transitions. As in the general 
Ginzburg-Landau approach, the Gross-Pitaevskii (GP) approach
is an expansion in powers of the order parameter field
up to fourth order. It works very well
close to the critical point, however, away from it, where the order
parameter is not small anymore, it may significantly deviate from the
correct result. This is not a problem in a homogeneous system with a uniform
order parameter, since we can restrict the theory to a regime
where the order parameter is small. In an inhomogeneous system, e.g., in a
trap, the order parameter varies in the system. Therefore, it is not 
sufficient, unless the system is very dilute,
to assume that the order parameter is small in some spatial region because it
can be large in another region of the system. The interparticle
interactions in the GP approach are approximated by a hard core two-body
potential with contribution only from the s wave scattering length. This 
approximation is quite satisfactory for the extremely dilute gases
satisfying the condition $na^{3}\ll 1$, where $n$ is density of particles
and $a$ is the scattering length. The physical implication of this condition
is that it is highly improbable for three or more particles to
interact with each other simultaneously. Therefore, it is clear that as the 
density of particles in the condensate increases,  the likelihood of the
three-body and higher order interactions will also increase,
making it necessary to go beyond the GP approach. Even though the
recent experiments on the BE in magnetic traps were based on dilute systems
of bosons with $na^{3}\approx 10^{-6}$, it is foreseeable that experiments can
be performed where the Bose gas is dense. This is already indicated by the
history of these experiments: the number of particles in the condensate has
increased by three orders of magnitude~\cite{ketterle} as compared to 
the early results. In this context it is interesting to study the condensate
including higher order interaction effects, expected in systems at higher
density. The purpose of this article is to apply a method, which goes beyond the
GP approach, to analyse an interacting Bose gas in a trap at arbitrary densities.

The rest of the paper consists of two parts: in the first (Sect. II) part we
discuss the  case of a dilute Bose gas using the GP approach. This includes a
mean-field theory based on the non-linear Schr\"odinger equation and the
Thomas-Fermi
approximation. In the second part (Sect. III) the Bose gas with hard-core
interaction is defined as a functional integral in a slave boson 
representation.
From the latter we derive an effective functional integral for the order
parameter field which describes the BE condensate. The new effective
functional integral, which also constitutes the main result of this work, is
valid for an arbitrary condensate density, and takes account of the 
three-body
and higher order effects in the interparticle interactions, at a finite
temperature. We apply this approach to the problem of BE condensation in a 
trap in Sec III A, and study the behavior of the condensate wavefunction as a
function of chemical potential (which controls the particle density) by means 
of a mean-field theory based on the Thomas-Fermi approximation again. 
However, like the GP approach, the limitation of the present work is its 
inability to account for the effects of the atoms outside of the condensate 
on the condensate itself. Such an extension will be the subject of a future
publication.

\section{Dilute Bose Gas}
The Bose gas, defined as a grand canonical ensemble of bosons,
can be described in second quantization, for instance, using a
functional integral representation \cite{N/O}. The fluctuations with respect to
time (i.e., quantum fluctuations)
have a gap $k_BT$ at non-zero temperature $T$ due to the Matsubara
frequencies $\omega_n=nk_BT$
($n=0,1,...$). The condensation of bosons is characterized
by a spontaneous breaking of a $U(1)$--symmetry (the phase degree of freedom of
the complex boson field). This implies a Goldstone mode which describes
gapless
fluctuations in space. Thus the latter fluctuations are relevant for
the condensation whereas the fluctuations with $\omega_n\ne0$ can be neglected
because of the gap $k_BT$. Thus it is sufficient to consider the
$\omega_0$--component of the quantum field $\Phi(x,\omega_0)\equiv\Phi_x$,
since we are only interested in static properties of the condensate near the
phase transition. This approximation has been used in the
discussion of the condensation of a Bose gas in translational invariant
systems \cite{P,G} as well as in a harmonic trap \cite{BP,DS,S,F}. 

As an introduction we present the GP approach of a grand
canonical Bose gas. The latter is defined by the partition function \cite{N/O}
\begin{equation}
Z=\int e^{-S_{GP}}\prod_xd\Phi_xd\Phi^*_x.
\end{equation}
The bosons in the condensate are described by the complex field $\Phi$ which
is controlled by the GP action of a dilute hard-core Bose gas
\cite{P,G} 
\begin{equation}
S_{GP}=\sum_{x,x'}\Phi_xt_{x,x'}\Phi^*_{x'}
-\sum_x(\mu|\Phi_x|^2-{u\over2}|\Phi_x|^4).
\label{gpa}
\end{equation}
Since this model is based on a Bose gas with hard-core interaction \cite{HU},
only three independent parameters enter: the chemical potential $\mu$, the
scattering length
of the hard-core interaction $a$ and the mass of the bosons. The coupling
constant $u$ is proportional to the scattering length \cite{HU}, whereas the
mass enters into the hopping rate
$\tau=\sum_xt_{x,x'}<0$ which has the same energy scale as $\mu$.
Formally, the action (2) is defined on a lattice with lattice constant $a$.
This is a good approximation of the hard-core gas with scattering length $a$
if one is only interested in length scales large compared to $a$. This is
the case in the experiments because the typical size of the condensate is
$5\times10^{-4}$ cm whereas $a\approx 5\times10^{-7}$ cm \cite{DS}.
Usually the hopping term $\sum_{x,x'}\Phi_xt_{x,x'}\Phi^*_{x'}$ is replaced
by the continuum approximation $\Phi_x\tau(1+(1/6a^2)\nabla^2)\Phi^*_x$ and
the sum by a formal integral, for simplicity.

The magnetic trap can be modeled by introducing a
confining potential $V_x$, for instance, a harmonic potential.
It is convenient to use a dimensionless expression for the action in (2)
as given in Refs. \cite{BP,DS}.
The parameters of a gas with about 5000 $^{87}$Rb atoms, studied in the
experiment by Anderson et al. \cite{A}, has been estimated in Ref. \cite{DS}: 
The shape of the trap is anisotropic with the potential
$V_x=x_1^2+x_2^2+8x_3^2-\mu_0$. The coupling
constant of the interaction of the $^{87}$Rb-atoms is
$u\approx 813$ and the effective chemical potential $\mu_0-\tau\approx16.3$.

\subsection{Mean-Field Theory}
The properties of the condensate (e.g., the density)
can be evaluated using the saddle point
approximation of the action $S_{GP}$. This is
equivalent to the approximation which neglects fluctuations of $\Phi_x$
(classical field approximation). The classical field $\Phi_x$ is a
solution of the non-linear Schr\"odinger equation
\begin{equation}
((\tau/6a^2)\nabla^2+\tau+V_x+u|\Phi_x|^2)\Phi_x=0.
\end{equation}
This nonlinear differential equation is complicated and generally one
has to resort to numerical methods to obtain its exact solutions. However, 
in order to understand its behavior in the high-density limit one can
neglect 
the kinetic energy term because, in that case, the nonlinear term of the
equation is dominant. This is
known as the Thomas-Fermi approximation. 
In a translational invariant system (i.e., $V_x=-\mu_0$)
the Thomas-Fermi approximation gives a linear behavior for the condensate
density as a function of $\mu_0$
\begin{equation}
\rho_c\propto|\Phi_x|^2=(1/u)(\mu_0-\tau)\Theta(\mu_0-\tau),
\end{equation}
where $\Theta$ is the Heaviside step function.
The condensate density increases ad infinitum upon increasing $\mu_0$.
This behavior is in disagreement
with the depletion of the condensate expected at higher total densities.
It also contradicts the fact that the hard-core potential limits the
density of the Bose gas.  The reason behind this behavior is that in the
GP approach the hard-core condition is implemented by a two-body delta
function potential. However, in the high-density limit when a large
number of particles are close to each other, this potential does not
provide a strong enough repulsion needed for the strict imposition of
the hard-core condition. In other word, corrections due to three-body
and higher order effects will become equally important.
This reflects that the GP approach is realistic only
for a dilute Bose system: The linear behavior is correct near the critical
point where $|\Phi|^2\sim0$ but the slope of the density of the condensate
is less than linear as one goes away from the critical point. And finally it 
decreases, indicating the depletion of the condensate at higher total 
densities.
By the same argument one can also obtain the inhomogeneous condensate
density for the case of a high-density Bose gas composed of alkali atoms in
a trap confined by a potential $V_x$~\cite{BP}
\begin{equation}
\rho_x\propto|\Phi_x|^2=-(1/u)(V_x+\tau)\Theta(-V_x-\tau).
\end{equation}

 One may try to correct the high-density behavior of GP approach by
including terms such as $|\Phi_x|^6$, $|\Phi_x|^8$,$\ldots$ in the GP
action (Eq.(\ref{gpa})) to account for the three-body and higher order 
interactions terms. 
But our feeling is that one will need to go to very high orders in this 
expansion to obtain the correct limiting behavior. However, in this work we 
propose an alternative 
approach which restores the correct high-density density behavior of
an inhomogeneous Bose gas by imposing strict hard-core condition by
adopting a slave boson representation. This approach is an extension
of a similar approach formulated for the case of a homogeneous Bose
gas earlier, by one of us\cite{Z}.  In the following section we will develop 
and apply the aforementioned slave boson based approach to study the 
hard-core Bose gas in a harmonic trap.

\section{Slave Boson Approach}
The slave boson representation was originally developed
for fermion systems  (e.g., Hubbard model) \cite{SBA}. 
The advantage of using a slave boson representation in case of strongly 
interacting electrons is that one can account for many effects of
strong correlations at the mean-field level of this representation~\cite{SBA}.
For the Bose gas, in contrast to the Hubbard model of the fermion gas, the
dilute limit describes already interesting physics such as BE condensation.
The dilute limit of a Bose gas can be described quite adequately by traditional
approaches such as the GP approach, which takes into account only the 
two particle interactions. However, it is intuitively obvious that as the 
density of particles in the BE
condensate increases, the nature of interparticle interactions will 
become more and more complex, and one will need to take even three-body
and higher effects into account. 
It was demonstrated by one of us that the slave
boson representation allows one to describe the dynamics of a Bose gas
at arbitrary densities~\cite{Z}, which is what we review next.
For the case of bosons, the slave boson representation is even easier to
formulate because
there are only two states per site in a hard-core system: a site is either
empty (represented by a complex field $e_x$) or occupied by a single boson
(represented by a complex field $b_x$). A hopping process of a boson appears
as an exchange of an empty site with a singly occupied site. Following the
standard arguments \cite{SBA}, this picture
can be translated into a slave boson action $S_{s.b.}$ of the form
\begin{equation}
\sum_{x}[\sum_{x'}b^*_xe_xt_{x,x'}b_{x'}e^*_{x'}+V_x b^*_xb_x
+i\lambda_x(e^*_xe_x+b^*_xb_x-1)].
\end{equation}
Here again we have neglected fluctuations with respect to
time because they have a gap $k_BT$ at non-zero temperature $T$ as discussed in
Sect. II.
The field $\lambda_x$ enforces the constraint $e^*_xe_x+b^*_xb_x=1$ which
guarantees that a site is either empty or singly occupied. This becomes
clear if we consider the partition function where we integrate over all
fields
\begin{equation}
Z=\int e^{-S_{s.b.}}\prod_xd\lambda_xdb_xdb^*_xde_xde^*_x.
\end{equation}
The $\lambda$--field creates a Dirac delta function for the constraint.
The slave boson fields $e_x$ and $b_x$ can be combined to a collective field
$b^*_xe_x\to \Phi_x$. Then
the constraint field $\lambda$ and the slave boson fields can be integrated
out which leads finally to the action for the collective field $\Phi$. This
was demonstrated in detail in Ref.\cite{Z}. Here we only present the results:
The new partition function reads
\begin{equation}
Z=\int e^{-S_b-S_1}\prod_xd\Phi_xd\Phi^*_x
\label{eq-func}
\end{equation}
with hopping (``kinetic'') term
\begin{eqnarray}
S_b & = & \sum_{x,x'}\Phi_x(1-t)^{-1}_{x,x'}\Phi^*_{x'}
\approx\sum_x(1-\tau)^{-1}|\Phi_x|^2
\nonumber\\
& &+{\tau\over6a^2(1-\tau)^2}\Phi_x(\nabla^2\Phi^*)_x
\end{eqnarray}
and the potential term
\begin{eqnarray}
S_1 & = & -\sum_x\ln\Big(e^{-V_x/2-1/4}\int d\varphi_x e^{-\varphi_x^2}
\nonumber\\
& & \times{sinh(\sqrt{(\varphi_x-V_x/2)^2
+|\Phi_x|^2})\over\sqrt{(\varphi_x-V_x/2)^2+|\Phi_x|^2}}\Big).
\end{eqnarray}
$S_b+S_1$ is a generalization of the GP action to systems
with arbitrary density. It agrees with Eq. (2) in the dilute regime after
expanding $S_1$ in powers of $|\Phi_x|^2$ up to second order:
$S_1=a_1|\Phi_x|^2+a_2|\Phi_x|^4+o(|\Phi|^5)$
can be compared with the potential of the dimensionless GP
energy. For instance, we find for $V_x=0$ the coefficients $a_1=-1/6$ and
$a_2=1/180$. In general, both coefficients of the expansion depend on $V_x$.
That means the
composite field $\Phi$ of the slave boson approach must be rescaled by
$(813/2a_2)^{1/4}$ in order to get the equivalent of the field $\Phi$ of the
GP approach. Furthermore, there is a renormalized chemical
potential $(1-\tau)^{-1}-1/6$ instead of $\tau$ in (2).
$\tau$ can be fixed by comparing the slave boson result and the
GP result in the vicinity of a vanishing condensate.
We found with the above mentioned parameters $\tau\approx -5.5$. 
Going beyond the dilute regime the effective
potential of the slave boson representation deviates significantly from the
GP potential: it grows only linearly for large $|\Phi_x|$ in
contrast to the $|\Phi_x|^4$-behavior of the GP case.
Therefore, the
'confinement' of the condensate is much weaker and allows a destruction
of the latter by fluctuations. The strong potential in the GP
case makes the condensate very robust against fluctuations with
sufficiently large $|\Phi_x|$.

An interesting quantity for the characterization of the condensate is the
momentum distribution of condensate atoms~\cite{BP}. For the case of a 
inhomogeneous
condensate, the condensate wave function obviously is not an eigenfunction
of the momentum. However, we do expect the momentum distribution to be
sharply peaked around zero momentum and its shape to be determined both by 
the trap parameters and the interparticle interactions\cite{A}.
In the continuum limit that we are considering here, the momentum 
distribution can be evaluated from the 
of the field $\Phi$ as \cite{BP}
\begin{equation}
\langle\Phi_{k}\Phi_{-k}^*\rangle=\sum_{x,x'}e^{ik\cdot(x-x')}
\langle\Phi_{x}\Phi_{x'}^*\rangle,
\label{corr}
\end{equation}
where the average is taken with respect to the effective action $\langle
...\rangle=Z^{-1}\int ... e^{-S_b-S_1}\prod_xd\Phi_x d\Phi_x^*$.
In order to calculate this quantity we use a saddle point
approximation for the functional integral described in the next section. 

\subsection{Mean-Field Theory}
The total density and the density of the condensate in the trap can be
calculated again from the saddle point approximation. For the 
present case we feel that it is sufficient
because thermal fluctuations will not be important at the small 
temperatures at which the BE condensation occurs in these
systems. However, if one 
needed to account for thermal fluctuations, one could do so by studying
the deviations around the saddle point in the functional integral of
Eq.(\ref{eq-func})~\cite{N/O}.
That means we have to look for a solution
$\Phi_x$ which minimizes $S_b+S_1$. This problem is analogous to the
minimization of the GP energy, discussed above.
Instead of the  non-linear Schr\"odinger equation Eq. (3),
we have a generalization of this equation for the slave boson approach
\begin{equation}
[(\tau/6a^2)\nabla^2+(1-\tau)]\Phi_x+(1-\tau)^2{\partial S_1\over\partial
\Phi^*_x}=0.
\label{SPE}
\end{equation}
This non-linear differential equation is even more complicated than the
non-linear Schr\"odinger equation. Therefore, we apply again the
Thomas-Fermi approximation
\begin{equation}
\Phi_x+ (1-\tau){\partial S_1\over\partial\Phi^*_x}=0.
\label{TFA}
\end{equation}
$\partial S_1/\partial\Phi^*_x$ contains higher order terms (three-body
interactions etc.) which are important for the dense regime. However,
inclusion of the kinetic energy term will only be necessary if more
complicated structures (e.g., vortex states \cite{DS}) are considered.
Since the accuracy of the Thomas-Fermi approximation increases
with increasing density, it is particularly suitable for our case where we
are interested in the high-density regime.

The total density of bosons can be evaluated from $\partial\ln Z/\partial
\mu_0$. This quantity is measured with respect to a
lattice Bose gas with lattice spacing $a$. 
The maximal density is $n=1$, where we have one boson per site (i.e., one
boson in a volume element $a^3$).
In the trap the maximum is at the center and it decreases monotonically with
$|x|$ (cf. Fig. 1).
\begin{figure}
\leavevmode\centering\psfig{file=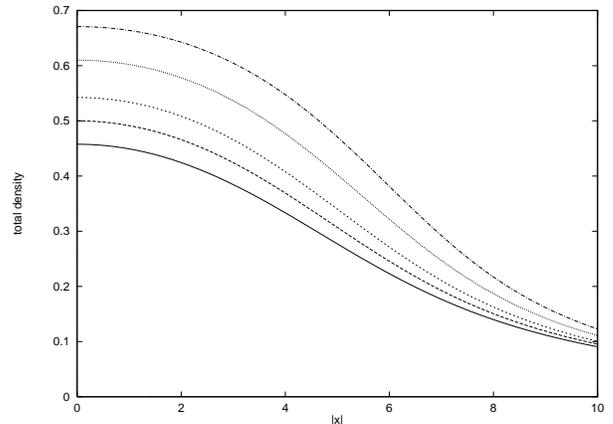,width=3.2 in}
\caption{Total density of bosons in a harmonic trap as a function of the
distance $|x|$ from the center of the trap for the chemical potential
$\mu_0=-0.6,0,0.6,1.6,2.6$. The total density at a given distance $|x|$
increases as the chemical potential increases. This result assumes the
equilibrium of the grand canonical ensemble of bosons, a condition which
may be violated in the experiment.}
\end{figure}

Solutions for an isotropic trap are presented for different values of $\mu_0$.
In reality, of course, the number of bosons $N$ is measured and $\mu_0$
must be determined selfconsistently for $N$ while
solving Eq.(\ref{SPE}).

\begin{figure}
\leavevmode\centering\psfig{file=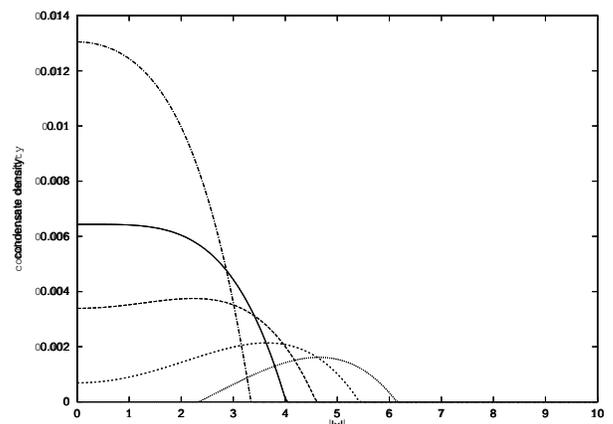,width=3.2 in}
\caption{Density of the condensate in the trap, normalized by the volume,
for the $\mu_0$ values of Fig.1. The surface of the condensate grows
with increasing $\mu_0$, and the condensate for the lowest density has
already the parabolic shape found in the GP approach.}
\end{figure}

However, in this case, we have used $\mu_0$ as a
parameter
which controls the number $N$, and hence, the density of atoms in the trap
and assigned it different values so as to solve 
Eq.(\ref{TFA}) for different values of particle density.
Solutions of Eq. (\ref{TFA}) for the density of the condensate are given 
in Fig.2.
The maximum of the condensate appears always at positions in the trap
where the total density is $0.5$. This reflects the duality of bosons and
holes in the hard-core Bose gas on the lattice.
If the total density is less than $0.5$
everywhere in the trap, the density of the condensate decays monotonically
with $|x|$. 
This result demonstrates that the condensate is depleted at the
center of the trap if the total density of bosons is larger than $0.5$.
Consequently, it is difficult to create an extended condensate in a
potential where the bosons are concentrated at the center with high density.
However, known experiments on a Bose gases are far away from such high
densities.

The correlation function of the order parameter field $\Phi_x=|\Phi_x|
e^{i\varphi_x}$ can be approximated by neglecting the fluctuations
of $|\Phi_x|$ as
\begin{equation}
\langle\Phi_{x}\Phi_{x'}^*\rangle
\approx {\bar\Phi}_x{\bar\Phi}_{x'}^*
\langle e^{i\varphi_x-i\varphi_{x'}}\rangle,
\end{equation}
where ${\bar\Phi}_x$ is a solution of Eq.(\ref{TFA}). 
(The global phase of ${\bar\Phi}_x$, of course, is not determined by 
Eqs.(\ref{SPE}) or (\ref{TFA}) due to symmetry.)
The phase coherence of the fluctuating phase $\varphi_x$ may be larger than
the size of the condensate because of the off-diagonal long range order in a
three-dimensional Bose gas \cite{popov}.
Therefore, the momentum distribution, defined by Eq.(\ref{corr}) and the
continuum limit can be approximated by \cite{BP}
\begin{equation}
f(k)=\Big|\int d^3x e^{ik\cdot x}{\bar\Phi}_x\Big|^2.
\label{mdist}
\end{equation}
The momentum distribution of the condensate atoms obtained using
Eq.(\ref{mdist}), based on the solution of Eq.(\ref{TFA}) is plotted in 
Fig.3 for several values of the chemical
potential $\mu_0$. 

\begin{figure}
\leavevmode\centering\psfig{file=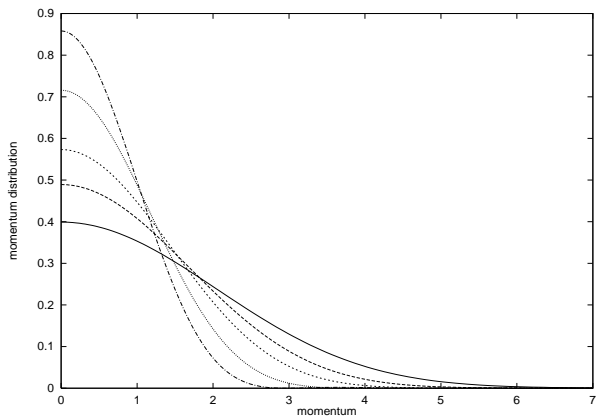,width=3.2 in}
\caption{Momentum distribution of condensate particles in the trap for
the $\mu_0$ values of Fig.1. The sharpness of the distribution increases
with increasing $\mu_0$}
\end{figure}

Upon inspecting Fig.3, one immediately observes 
that the momentum distribution becomes even sharper due to depletion.
This is a consequence of spreading of the condensate due to depletion which
supports small momenta.

\section{conclusions}
In conclusion, we have studied the condensation of a three-dimensional 
high-density Bose gas in a harmonic trap. We demonstrated the unphysical 
nature of the solutions that one obtains if one applies the traditional
approach of Gross-Pitaevskii to study such a system. We identified that
the reason behind this behavior was that the GP action ignores the  
three-body and the higher order interactions which become important
at high densities. In this work we proposed an alternative approach
based on the slave boson representation, which accounts for these complex
interactions at high densities by satisfying the hard-core condition
strictly. Our approach leads to solutions which are well behaved at
high densities and predicts depletion of the condensate in the regions
of high densities (mainly center of the trap) which one would expect on 
intuitive grounds. We also
study the momentum distribution of the atoms in the trap and observe a
narrowing of the momentum distribution in the high-density limit. 
This approach is expected to be of little application for the low-density
condensates which are being created in the experiments presently, but
we hope that it can be tested in future when experimentalists may be
able to realize a dense Bose gas in a trap. We also plan to apply this approach
to study the structure of vortices in a high-density inhomogeneous Bose gas,
results of which will be presented in a future publication.

Acknowledgement: We would like to thank Professor L.P. Pitaevskii for a
stimulating discussion.

\end{document}